\documentclass[12pt]{iopart}
\usepackage{iopams}
\usepackage{graphicx}
\usepackage{amssymb}
\usepackage{epstopdf}
\usepackage{subfigure}
\usepackage{float}

\usepackage{color} 

\usepackage{iopams}

\begin{document}
\title[Role of dimensionality in preferential attachment growth in the Bianconi-Barab\'asi model]{Role of dimensionality in preferential attachment growth in the Bianconi-Barab\'asi model}
\author{Thiago C. Nunes$^1$, Samurai Brito$^{2}$, Luciano R. da Silva$^{1,3}$ and Constantino Tsallis$^{3,4}$}
\address{$^1$ Universidade Federal do Rio Grande do Norte, Departamento de F\'isica Te\'orica e Experimental, Natal-RN, 59078-900, Brazil.}
\address{$^2$ International Institute of Physics, Natal-RN, 59078-970, Brazil.}
\address{$^3$ National Institute of Science and Technology of Complex Systems, Brazil.}
\address{$^4$ Centro Brasileiro de Pesquisas F\'isicas, Rua Xavier Sigaud 150, 22290-180 Rio de Janeiro-RJ, Brazil; \\ Santa Fe Institute, 1399 Hyde Park Road, New Mexico 87501, USA; \\ Complexity Science Hub Vienna, Josefstaedter Strasse 39, A 1080 Vienna, Austria.}
\ead{thiagoccn@dfte.ufrn.br, samuraigab@gmail.com, luciano@dfte.ufrn.br, tsallis@cbpf.br}

\begin{abstract}
Scale-free networks are quite popular nowadays since many systems are well represented by such structures. In order to study these systems, several models were proposed. However, most of them do not take into account the node-to-node Euclidean distance, i.e., the geographical distance. In real networks, the distance between sites can be very relevant, e.g., those cases where it is intended to minimize costs. 
Within this scenario we studied the role of dimensionality $d$ in the Bianconi-Barab\'asi model with a preferential attachment growth involving Euclidean distances. The preferential attachment in this model follows the rule $\Pi_{i} \propto \eta_i k_i/r_{ij}^{\alpha_A}$ $(1 \leq i < j; \alpha_A \geq 0)$, where $\eta_i$ characterizes the fitness of the $i$-th site and is randomly chosen within the $(0,1]$ interval. We verified that the degree distribution $P(k)$ for dimensions $d=1,2,3,4$ are well fitted by $P(k) \propto e_q^{-k/\kappa}$, where $e_q^{-k/\kappa}$ is the $q$-exponential function naturally emerging within nonextensive statistical mechanics. We determine the index $q$ and $\kappa$ as functions of the quantities $\alpha_A$ and $d$, and numerically verify that both present a universal behavior with respect to the scaled variable $\alpha_A/d$. The same behavior also has been displayed by the dynamical $\beta$ exponent which characterizes the steadily growing number of links of a given site.
\end{abstract}


\vspace{2pc}
 
\submitto{\JSTAT}

\maketitle

\section{Introduction}
Scale-free networks emerge spontaneously in many systems, and due to their uncountable applications in different fields of knowledge, like physics, biology, economics, social sciences, among other areas, has become very popular nowadays \cite{strogatz2001,newman2003,barabasi1999,wattsstrogatz1998}. Many of these networks are characterized by an asymptotic power law degree distribution, instead of the usual exponential laws. There are some networks where the Euclidean distance between sites are very relevant and cannot be neglected, once it plays a key role in maximizing or minimizing costs. These networks are named \textit{geographic networks} \cite{Barthlemy2011, Bianconi2016}. Like examples, we have airlines networks, electrical power grid or even pipes that carry water to the home. Some models \cite{soares2005,meneses2006,thurnertsallis2005} in the literature take into account this aspect of the problem and reveals strong connection with nonextensive statistical mechanics \cite{tsallis1988,gellmann2004,tsallis2009}, based on the nonadditive entropy $S_q = k\frac{1-\sum_i p_i^{q}}{q - 1}$ ($q \in \Re$; $S_1 = S_{BG} \equiv -k \sum_i p_i\ln p_i$ where BG stands for Boltzmann-Gibbs).

This current generalization of the BG entropy has been widely successful in explain thermostatistical properties in complex systems in high-energy collisions at LHC/CERN (CMS, ALICE, ATLAS and LHCb detectors) and at RHIC/Brookhaven (PHENIX detector) \cite{adare2011,wong2013,marques2013,aaij2016}, cold atoms \cite{douglas2006}, dusty plasmas \cite{goree2008}, among others. 

The deep relationship between scale-free networks and $q$-statistics started being explored in 2005 \cite{soares2005}, where it was shown that it is possible to relate the $q$-exponential function to the scale-free networks through the connectivity distribution $P(k)$. A large variety of these networks exhibits a power law tail and their degree distribution has the form $P(k) \sim \frac{1}{(k_0 + k)^{\gamma}}$. Using the basic functional $S_q[P(k)] = k \frac{1-\int dk [P(k)]^{q}}{q-1}$ with the constraint $\left\langle k\right\rangle \equiv \int dk k P(k) =$ constant ($k$ being the degree of a generic site), it has  been shown that:

\begin{equation}
 P(k) = P(0)e_q^{-k/\kappa} = P(0)[1 + (q - 1)k/\kappa]^{\frac{1}{1-q}} \label{Eq1}
\end{equation}
for $q > 1$ and $k \rightarrow \infty$, $P(k) \sim k^{-\gamma}$ with $\gamma \equiv 1/(q - 1)$, thus revealing to be the generic degree distribution for a wide class of scale-free networks, with the $q$-exponential function being defined as: 
\begin{equation}
e_q^{z} \equiv [1 + (1 - q)z]^{\frac{1}{1-q}}\ (e_1^{z} = e^{z}), \label{Eq2}
\end{equation}
and its inverse function being defined as:
\begin{equation}
\ln_q(z) \equiv \frac{z^{1-q}-1}{1-q}, \forall (z,q). \label{Eq3}
\end{equation} 
\section{Methods}
In this paper we study the role of dimensionality in the Bianconi-Barab\'asi model \cite{bianconi2001} for $d=1,2,3,4$ dimensions (from now when we discuss about dimensions in this paper, $\forall d$, specifically $d = 1,2,3,4$). In order to study this, we included the term $r_{ij}$ in the preferential attachment rule, being $r_{ij}$ the Euclidean distance from site $i$ to the newly arrived site $j$. To generate a network by this model, firstly we choose the dimension $d$ of the system. To each new site that arrives at the network is attributed a position and a fitness parameter $\eta_i$, which is randomly chosen from a uniform distribution and belongs to the interval $\in (0,1]$. From now when we discuss about the variable $\eta_i$ in this paper, $\forall \eta$, specifically $\eta = 0.3,0.6, 0.9$ (typical values of $\eta$). To fix the position of the sites in the line, plane, space and so on, we proceed in the following way. 

We considered the center of mass as being the origin of the system.  The first site ($i=1$) is put in an arbitrary origin. Sequentially we added the second node ($i = 2$) at an Euclidean distance $r$ from the site $i = 1$ obeying the d-dimensional isotropic distribution:
 
\begin{equation}
p(r) = \frac{1}{r^{d + \alpha_G}}\ (\alpha_G > 0; \forall d), \label{Eq4}
\end{equation}
where $r \geq 1$ (in one dimension, $r = \left|x\right|$; in two dimensions, $r = \sqrt{x^2+y^2}$; in three dimensions $r = \sqrt{x^2+y^2+z^2}$, and so on); we assumed angular isotropy. From this step on, the center of mass of the system is always recalculated and then the new site ($i > 2$) is positioned at a distance $r$ from the current origin obeying  Eq. (\ref{Eq4}). The exponent $\alpha_G$ controls the network growth and the subindex G stands for \textit{growth}. The newly arrived site will then be connected to one of the pre-existing sites of the network through the preferential attachment rule given by: 

\begin{equation}
    \Pi_{i} = \frac{k_i \eta_i r_{ij}^{-\alpha_A}}{\sum_i k_i \eta_i r_{ij}^{-\alpha_A}} \in (0,1)\ (\alpha_A \geq 0), \label{Eq5}
\end{equation}
where $k_i$ is the connectivity of the $i^{th}$ pre-existing site; $\alpha_A$ controls the importance of the distance in the preferential attachment rule and the subindex A stands for \textit{attachment}. The previous procedure are used to include the third site, fourth site, and so on up to a given $N$ ($N$ is the total number of nodes).

The preferential attachment of this model makes the pre-existing sites to compete for links and three factors can influence the probability of the sites to receive the new connections: the sites degree, the Euclidean distance of this sites to the new site arriving in the network, and the fitness of these sites. The importance of the distance in the preferential attachment rule is less pronounced when $\alpha_A$ is close to zero and completely disappears for $\alpha_A = 0$. In this limit, we recover the Bianconi-Barab\'asi Model, which has topology but no metrics. It is important to say that the model studied here, although more general, does not contain the model studied by Brito et al. \cite{britosilvatsallis2016} as a particular case. Therefore, it is not possible to recover here the results found by these authors.

In the present paper, we will focus on the following main aspects:
\begin{itemize}
    \item The connectivity distribution $P(k)$ of the network and its dependence on $(\alpha_A,d)$.
    
    \item The average connectivity time evolution of the site, more precisely how $\left\langle k_i \right\rangle$ grows with the scaled time $t/t_i \ (t \geq t_i)$. Given that $k_i(t) \propto (t/t_i)^{\beta}$, we studied the dynamical $\beta$ exponent as a function of $(\alpha_A,d)$, as well as of the fitness $\eta$ of a arbitrarily chosen illustrative site.
\end{itemize}

\section{Results}
\begin{figure}[H]
\begin{center}
\includegraphics[scale = 0.45]{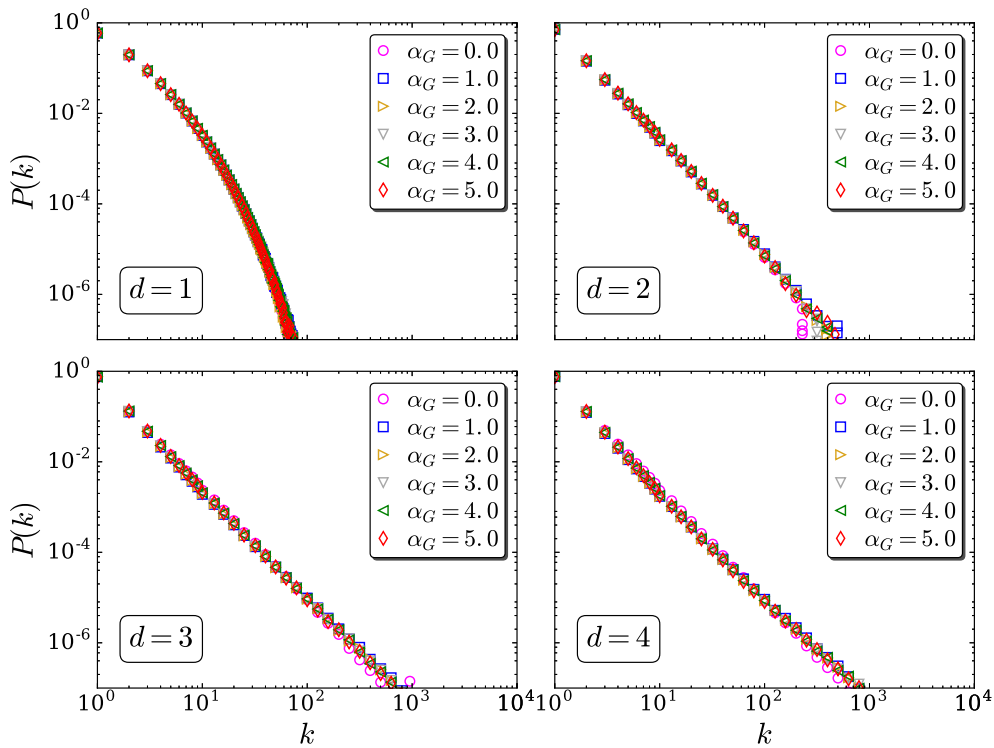}
\caption{\small Connectivity distribution for $d = 1, 2, 3, 4,$ $\alpha_A = 2.0$ and typical values for $\alpha_G$. The simulations have been run for $10^{3}$ samples of $N = 10^{5}$ sites each. We verified that $P(k)$ is independent from $\alpha_G \ (\forall d)$. Logarithmic binning was used whenever convenient.} 
\label{fig1}
\end{center}
\end{figure}

\begin{figure}[H]
\begin{center}
\includegraphics[scale = 0.45]{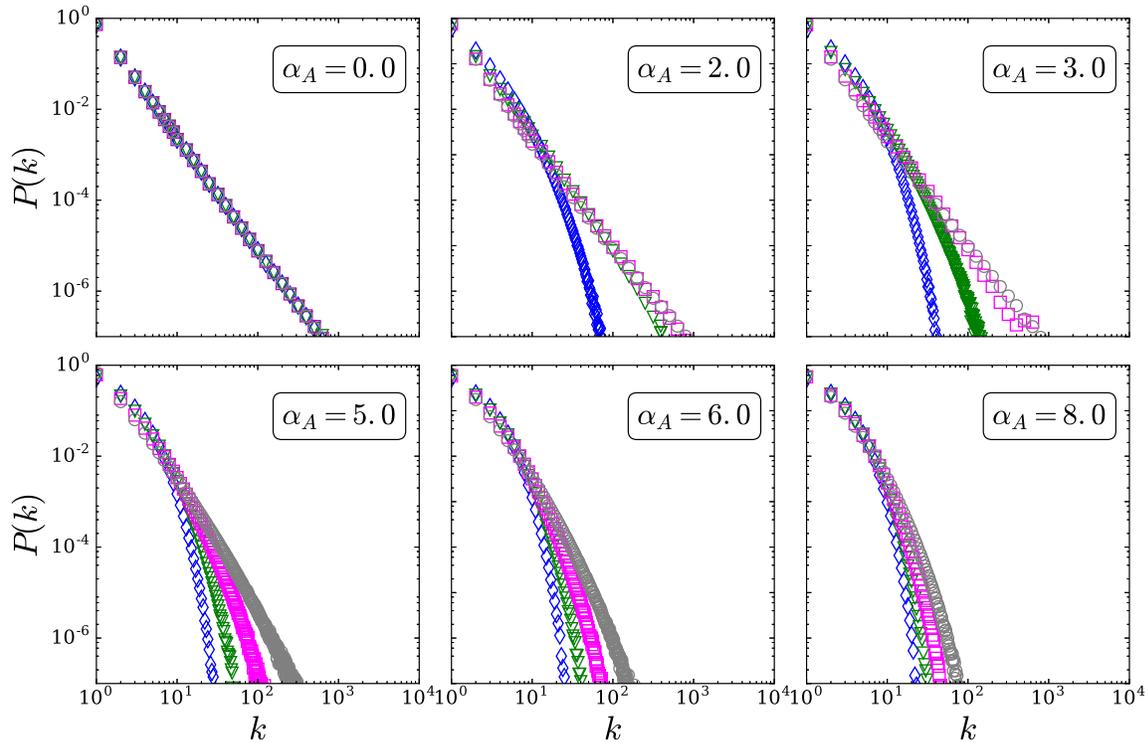}
\caption{\small Degree distribution for $d = 1$ (blue diamonds), $2$ (green triangles), $3$ (magenta squares), $4$ (grey circles), and typical values of $\alpha_A$ for fixed value of $\alpha_G=2$. The simulations have been run for $10^{3}$ samples of $N=10^{5}$ sites each. Logarithmic binning was used whenever convenient.} 
\label{fig2}
\end{center}
\end{figure}

Our numerical simulations for $P(k)$ indicated in the figure \ref{fig1} and \ref{fig2} have been performed, $\forall d$, by varying the $\alpha_G$ and $\alpha_A$ parameters. We have verified in all cases that $P(k)$ does not depend on $\alpha_G$ for any value of $\alpha_A$ and $d$ (see figure \ref{fig1}). Using this fact, we have arbitrarily fixed $\alpha_G = 2$ and numerically studied the influences of $\alpha_A$ and $d$ on $P(k)$. Similarly to the results found in  \cite{britosilvatsallis2016}, we have observed, in the present model, that $\alpha_A$ has a strong influence on $P(k)$ (see figure \ref{fig2}).  In this way, we numerically verified, for our model, that the connectivity distributions are very well fitted by the function $P(k) = P(0)e_q^{-k/\kappa}$ for all $\alpha_A$ and $d$, with $q > 1$ and $\kappa > 0$. However, when $\alpha_A = 0$, we have the Bianconi-Barab\'asi model, whose exact solution is known and given by $P(k) = k^{-\gamma}/\ln(k)$, with $\gamma = 2.255$ \cite{Ergn2002, Newman2010}. When $\alpha_A > 0$, no exact solution is known and our numerical results strongly suggest that this model has a degree distribution which is, in fact, $P(k) \propto e^{-k/\kappa}_{q}$.

For $\alpha_A \to \infty$ ($q \to 1$) the BG limit is reached and $P(k)$ tends to the standard exponential function. In this limit, independently of the system dimension, the network present typical connectivity between sites, that is a characteristic of classical random networks. This fact can be seen as a topological phase transition (from scale-free to random networks) associated with the $\alpha_A$ parameter.

\begin{figure}[H]
\begin{center}
\includegraphics[scale = 0.45]{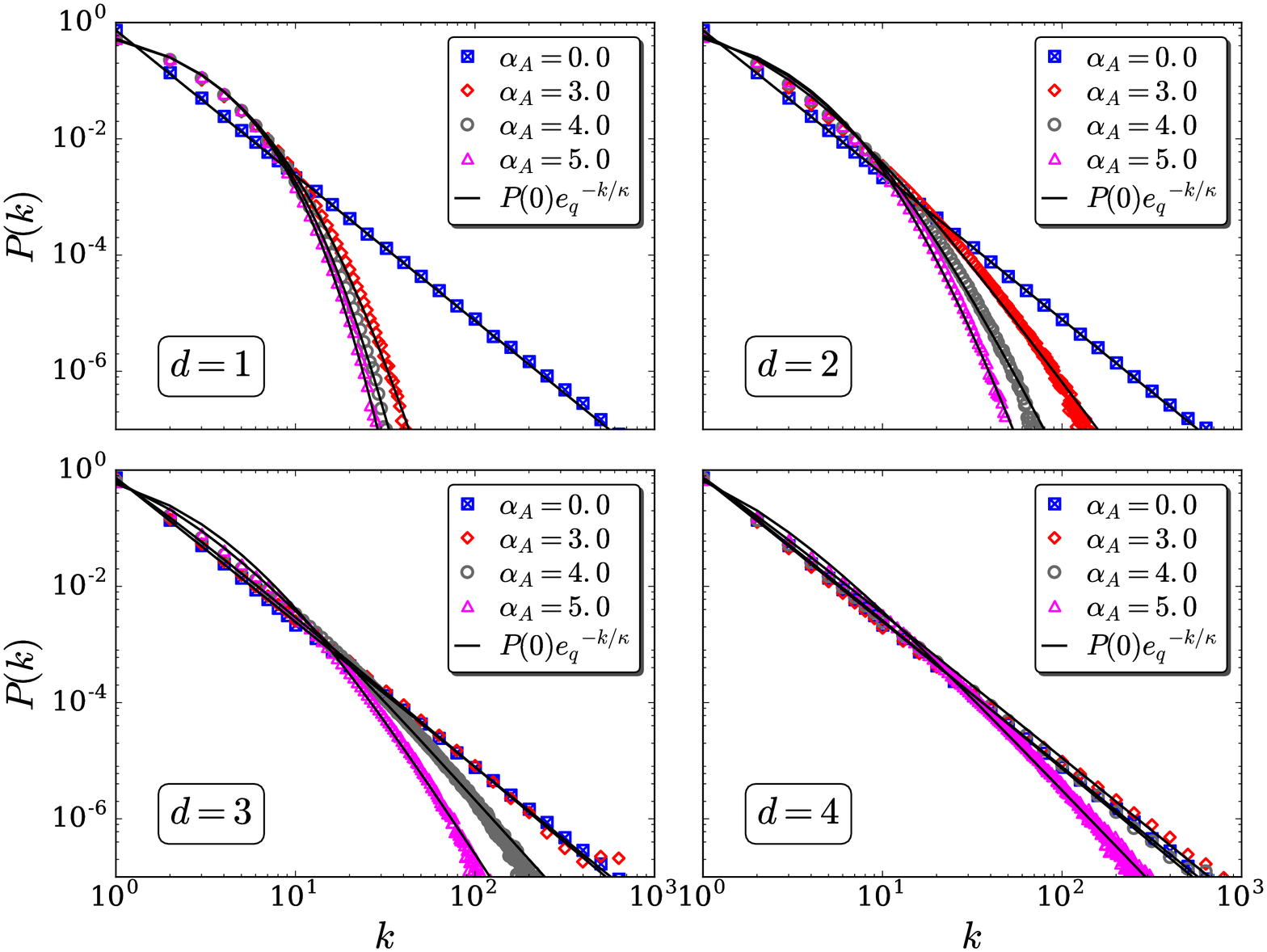}
\includegraphics[scale = 0.40]{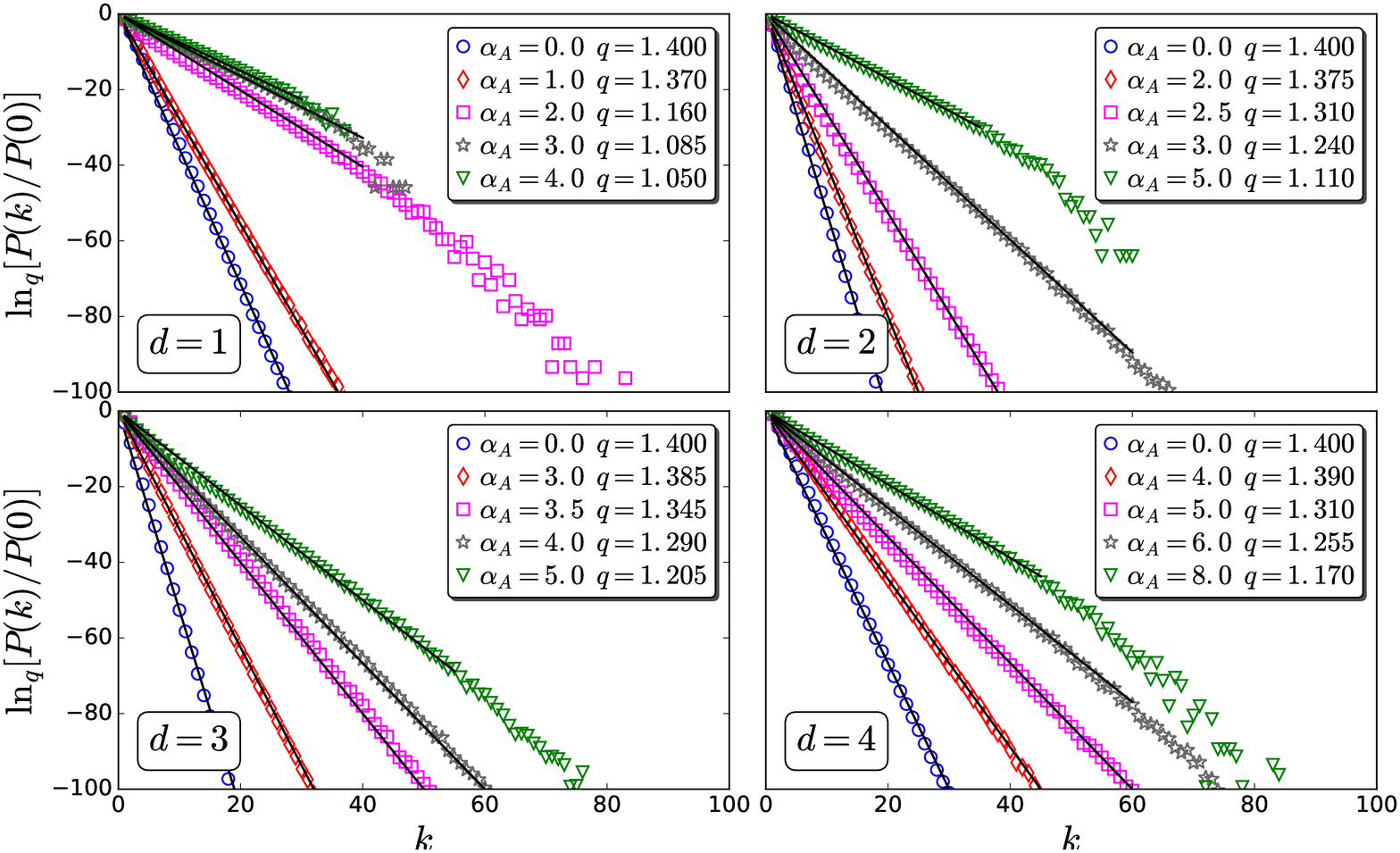}
\caption{\small Connectivity distributions for $d = 1, 2, 3, 4$, fixed $\alpha_G=2$ and typical values of $\alpha_A$. We can see that all curves of $P(k)$ are very well represented by $q$-exponential functions (defined in Eq. (\ref{Eq2})) in such way that $P(k) = P(0)e^{-k/\kappa}_{q}$. In the top figure we have log-log representation. In the bottom we have $\ln_q[P(k)/P(0)]$ versus $k$ representation. The numerical fluctuations for increasing $k$ will disappear in the thermodynamic limit $N \rightarrow \infty$. Logarithmic binning was used whenever convenient.} 
\label{fig3}
\end{center}
\end{figure}
\begin{figure}[H]
\subfigure {\includegraphics[scale = 0.40]{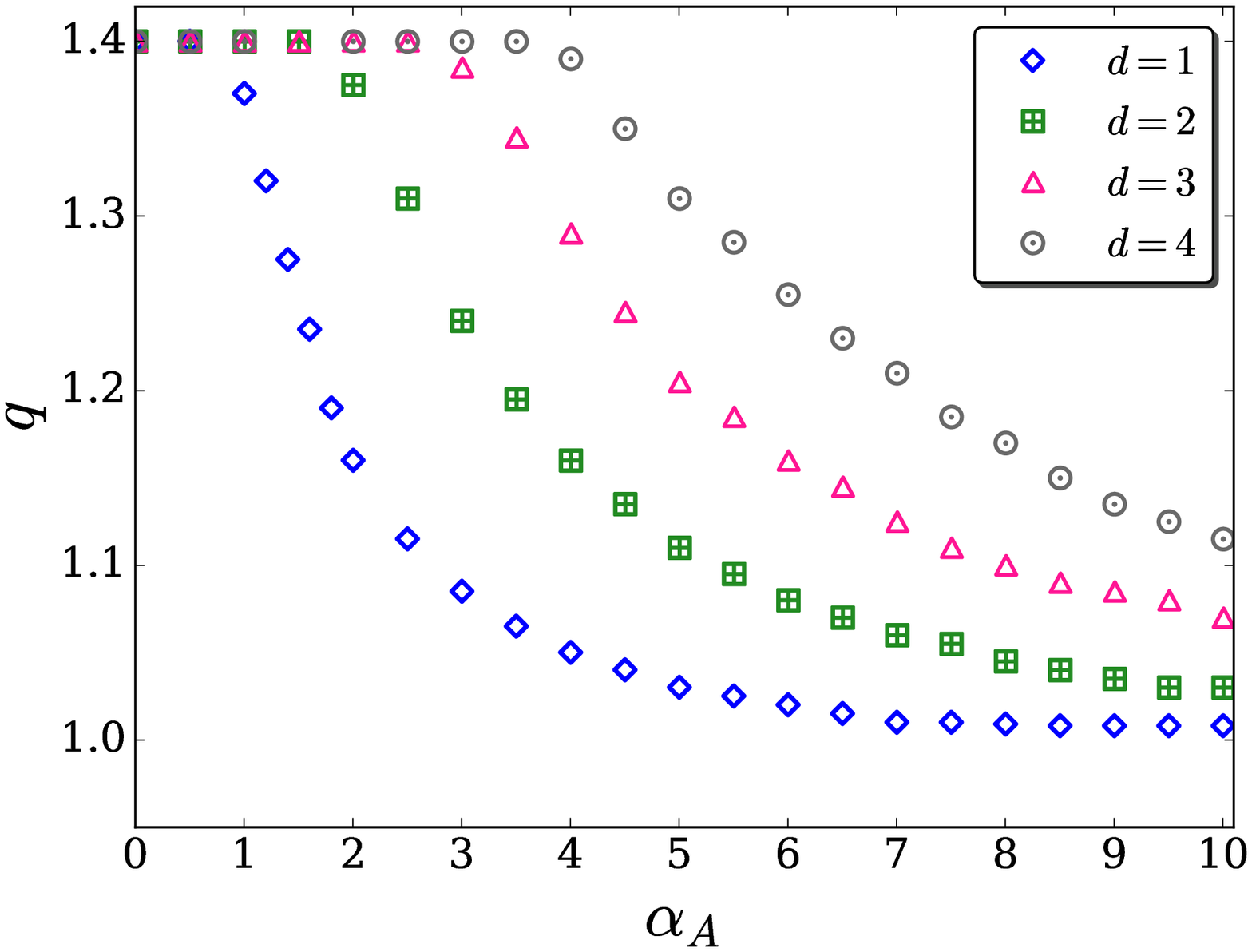}}
\subfigure {\includegraphics[scale = 0.40]{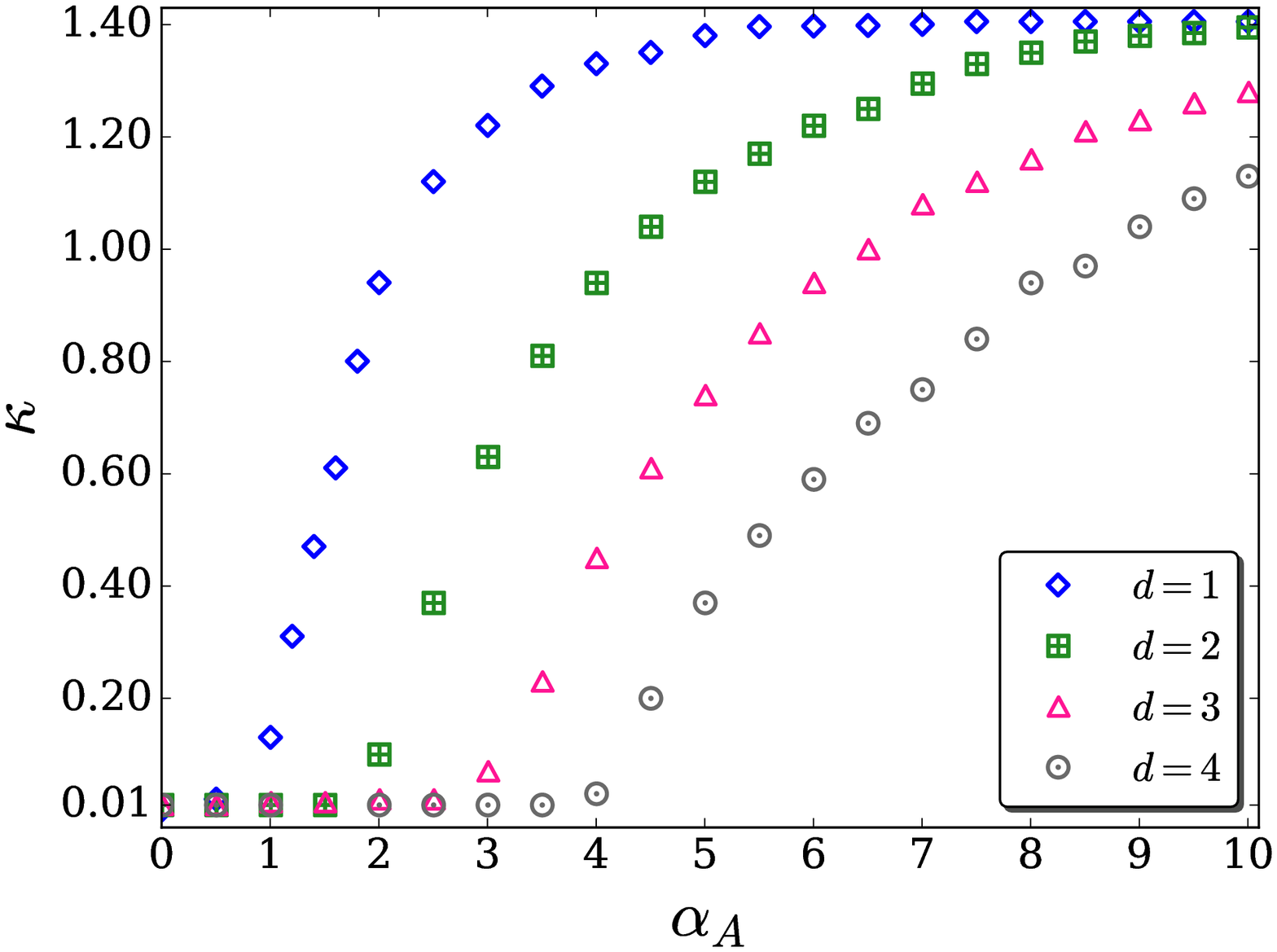}}
\caption{\small $q$ and $\kappa$ for $d = 1, 2, 3, 4$. We can see how $q$ and $\kappa$ vary with $\alpha_A$ and $d$. $q$ has an upper limit ($q=7/5$) and $\kappa$ has a lower limit ($\kappa=0.01$) regardless of the system dimension.} 
\label{fig4}
\end{figure}
\begin{figure}[H]
\subfigure {\includegraphics[scale = 0.32]{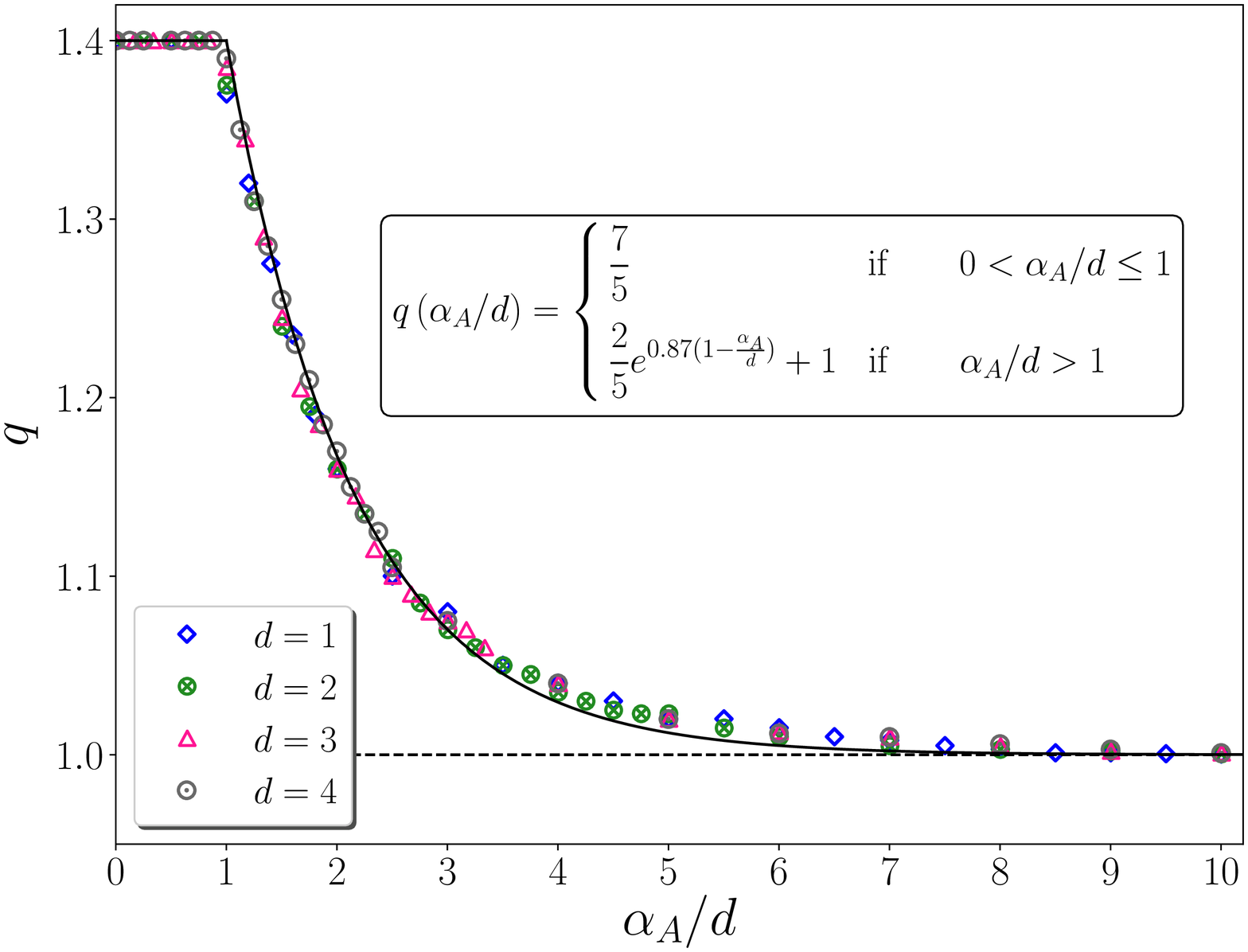}}
\subfigure {\includegraphics[scale = 0.32]{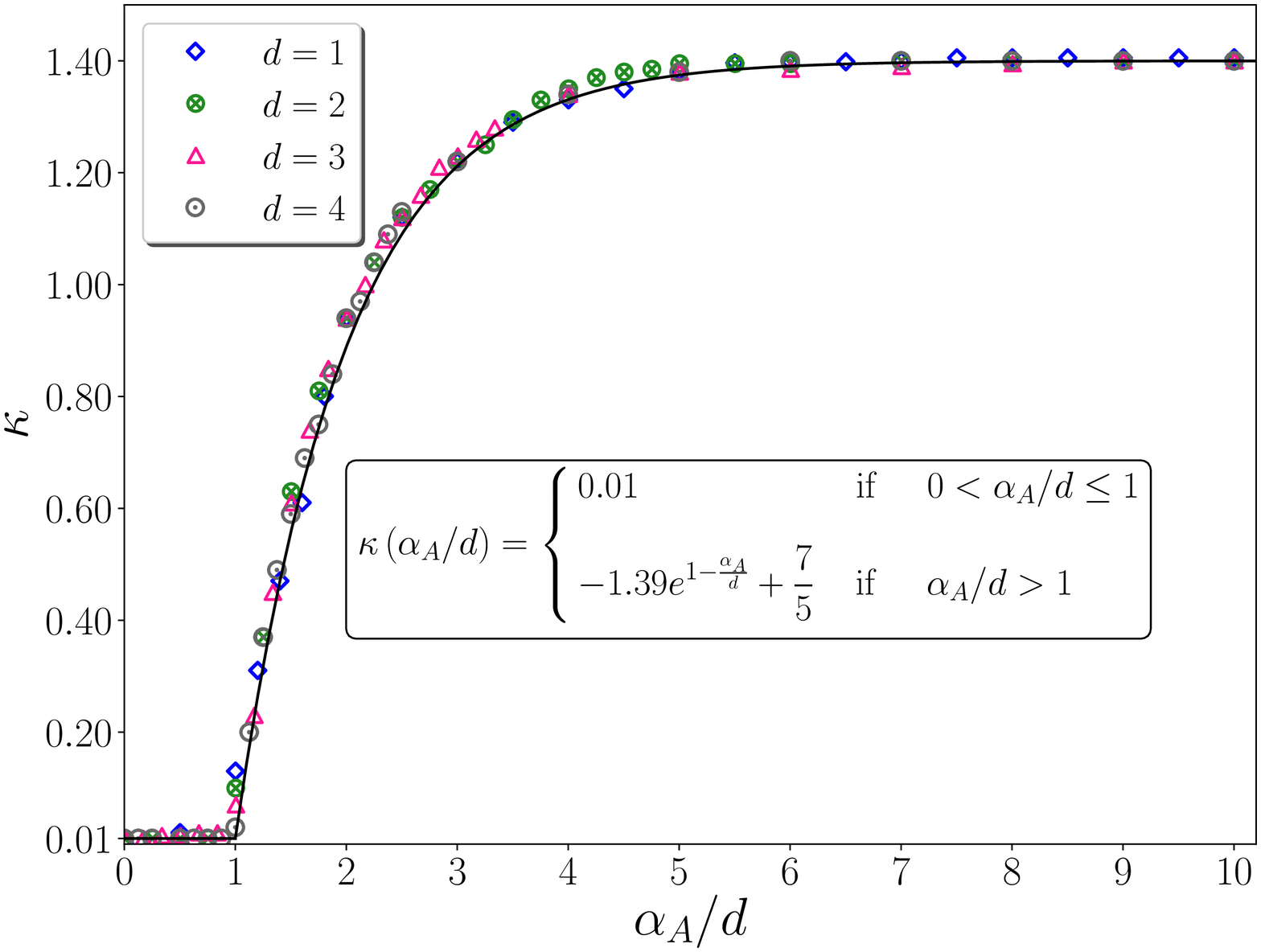}}
\caption{\small $q$ and $\kappa$ versus $\alpha_A/d$ (same data as in the figure \ref{fig5}). We see that $q = 7/5$ and $\kappa = 0.01$ for $0 < \alpha_A/d \leq 1$. Nearly exponential behavior gradually emerges for $\alpha_A/d > 1$ $(\forall d)$ and similarly for $\kappa$. These results exhibit the universality of both $q$ and $\kappa$.} 
\label{fig5}
\end{figure}

Our results showed, furthermore, that for each $\alpha_A$ values we have one corresponding $q$ and $\kappa$ value, where $q$ is the entropic index and $\kappa$ is the characteristic number of links (or effective `temperature') (see figure \ref{fig3}). Once $q$ and $\kappa$ were obtained from the fits, we further analyzed how these parameters vary with respect to $\alpha_A$ and $d$. We can see in the figure \ref{fig4} that there is a special value of $\alpha_A$ ($\alpha_A=d$) from which $q$ decreases with $\alpha_A$ while, in contrast, $\kappa$ increases. In the figure \ref{fig5} we can see that, although both indices $q$ and $\kappa$ depend on $\alpha_A$ and $d$, they present remarkably enough universal curves with respect to the scaled variable $\alpha_A/d$. When $0 \leq \alpha_A/d\leq 1$ the system presents long-range interactions and, except for $\alpha_A = 0$, it is characterized by $q=7/5$ and $\kappa = 0.01$. When $\alpha_A/d > 1$ the nearly exponential behavior gradually emerges for $q$ and $\kappa$. Both $q$ and $\kappa$ quickly approach to their corresponding BG limits ($q = 1$) for $\alpha_A/d \rightarrow \infty$. Moreover, the same exponential $e^{1-\alpha_A/d}$ appears in both expressions for $q$ and $\kappa$.

Consequently, the following nearly linear\footnote{In the same spirit, the linear relation between $q$ and $\kappa$, found by Brito et al. \cite{britosilvatsallis2016}, can be improved with the following nearly linear relation $\kappa \simeq 1.45 - 4.06(q-1)^{1.15}$.} relation can be established:

\begin{equation}
    \kappa \simeq 1.4 - 3.98(q-1)^{1.15}. \label{Eq6}
\end{equation}

\begin{figure}[H]
\begin{center}
\includegraphics[scale = 0.45]{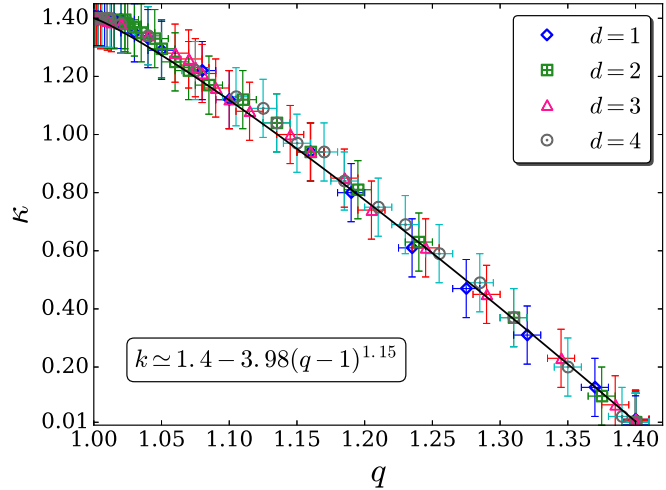}
\caption{\small $q$ and $\kappa$, for the present models $(d = 1, 2, 3, 4)$,  follow nearly linear relation given by $\kappa \simeq 1.4 - 3.98(q-1)^{1.15}$ (continuous straight line). The up most value of $q$ is $7/5$, yielding $\kappa = 0.01$ $(\forall d)$.} 
\label{fig6}
\end{center}
\end{figure}

In fact, this simple relation is numerically quite well satisfied as can be seen in the figure \ref{fig6}. Similarly, a nearly linear relation between $q$ and $\kappa$ was found in another model \cite{britosilvatsallis2016}. 
This fact expresses a frequent peculiarity of $q$-statistics. We can see, once again, that $\kappa$ is not a free parameter, but it is instead an intrinsic characteristic of the system.

\begin{figure}[H]
\begin{center}
\includegraphics[scale = 0.40]{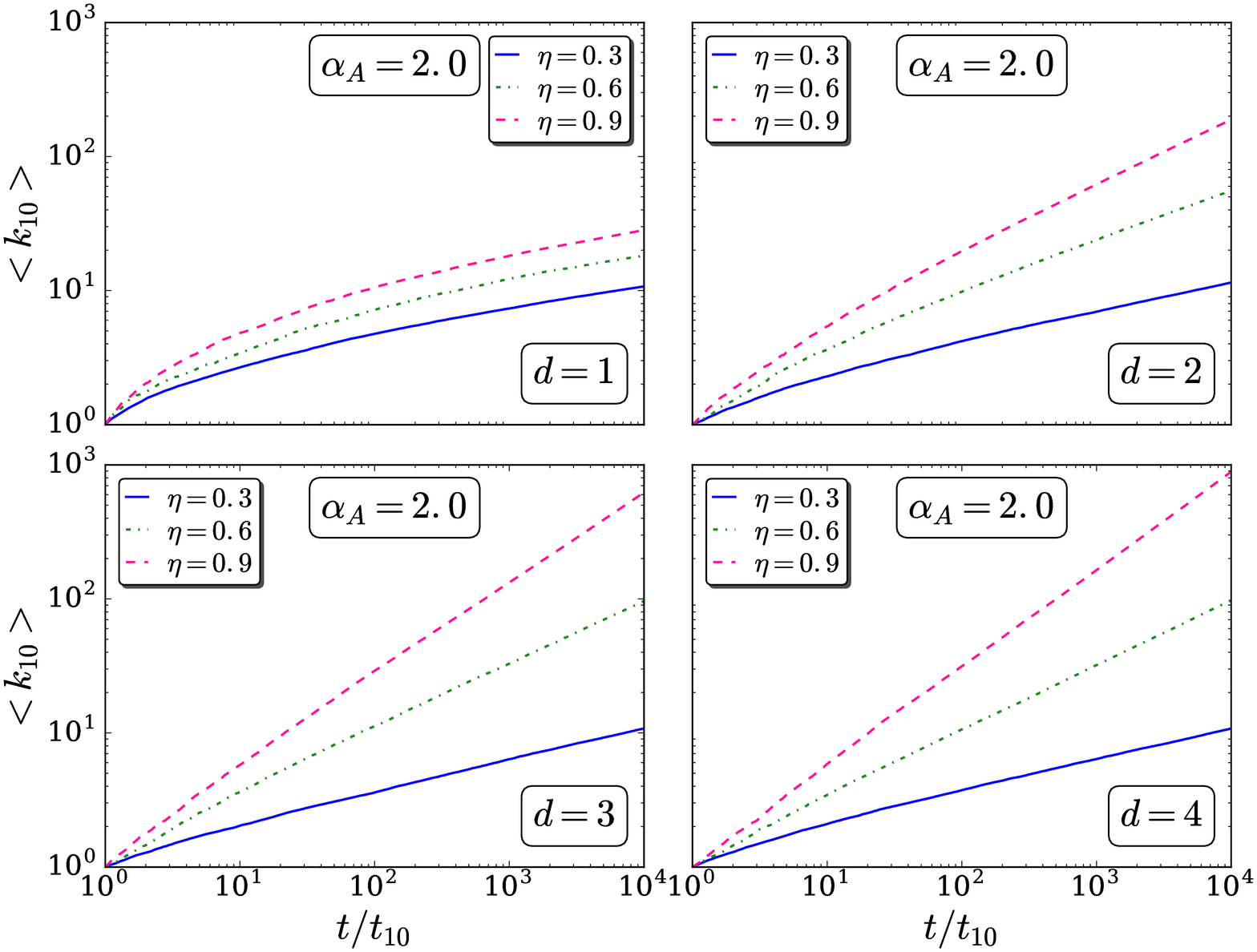}
\includegraphics[scale = 0.40]{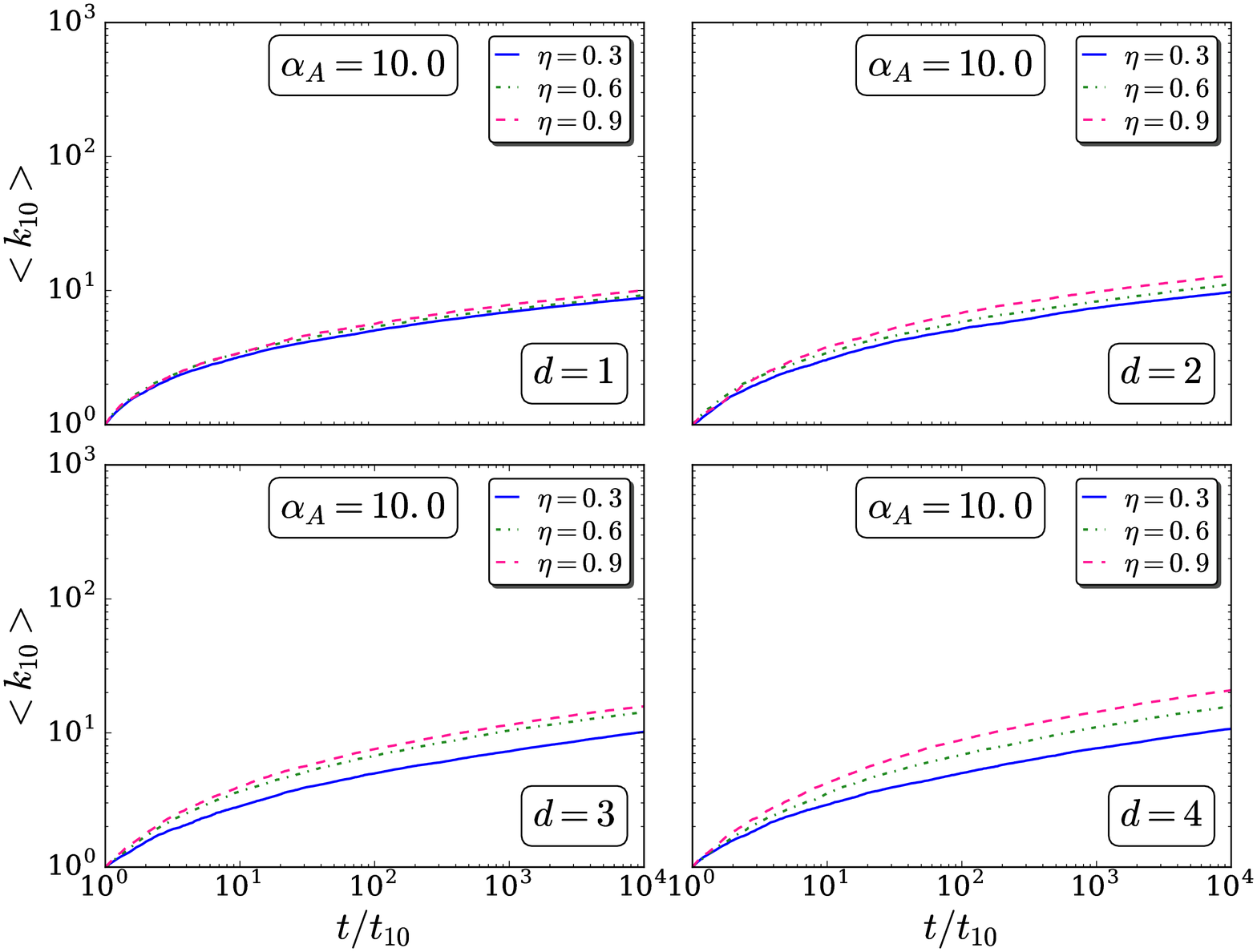}
\caption{\small Connectivity time evolution of the site $i=10$ for two $\alpha_A$ values ($\alpha_A=2$ and $\alpha_A=10$). In this figure, we show how the connectivity of this site evolve with $\eta$, $\alpha_A$ and $d$. We can see that the slope of the curves increases with $d$ and $\eta$, however, decreases one with $\alpha_A$.} 
\label{fig7}
\end{center}
\end{figure}

The connectivity time evolution $k_i(t)$ was calculated and it is indicated in the figure \ref{fig7}. We have chosen the site $i=10$ for our analysis, but the connectivity time evolution is independent of the choice of site $i$. We have shown that the connectivity time evolution varies only with $\alpha_A$, $\eta$ and $d$. Our data for $k_i(t)$ obey the function:

\begin{equation}
    k_i(t) \propto \left(\frac{t}{t_i}\right)^{\beta(\alpha_A, \eta_i, d)} t \geq t_i .\label{Eq7}
\end{equation} 
We can see that the $\beta$ exponent increases with $d$ and $\eta$, but decreases with the $\alpha_A$ parameter.

\begin{figure}[H]
\begin{center}
\subfigure {\includegraphics[scale = 0.37]{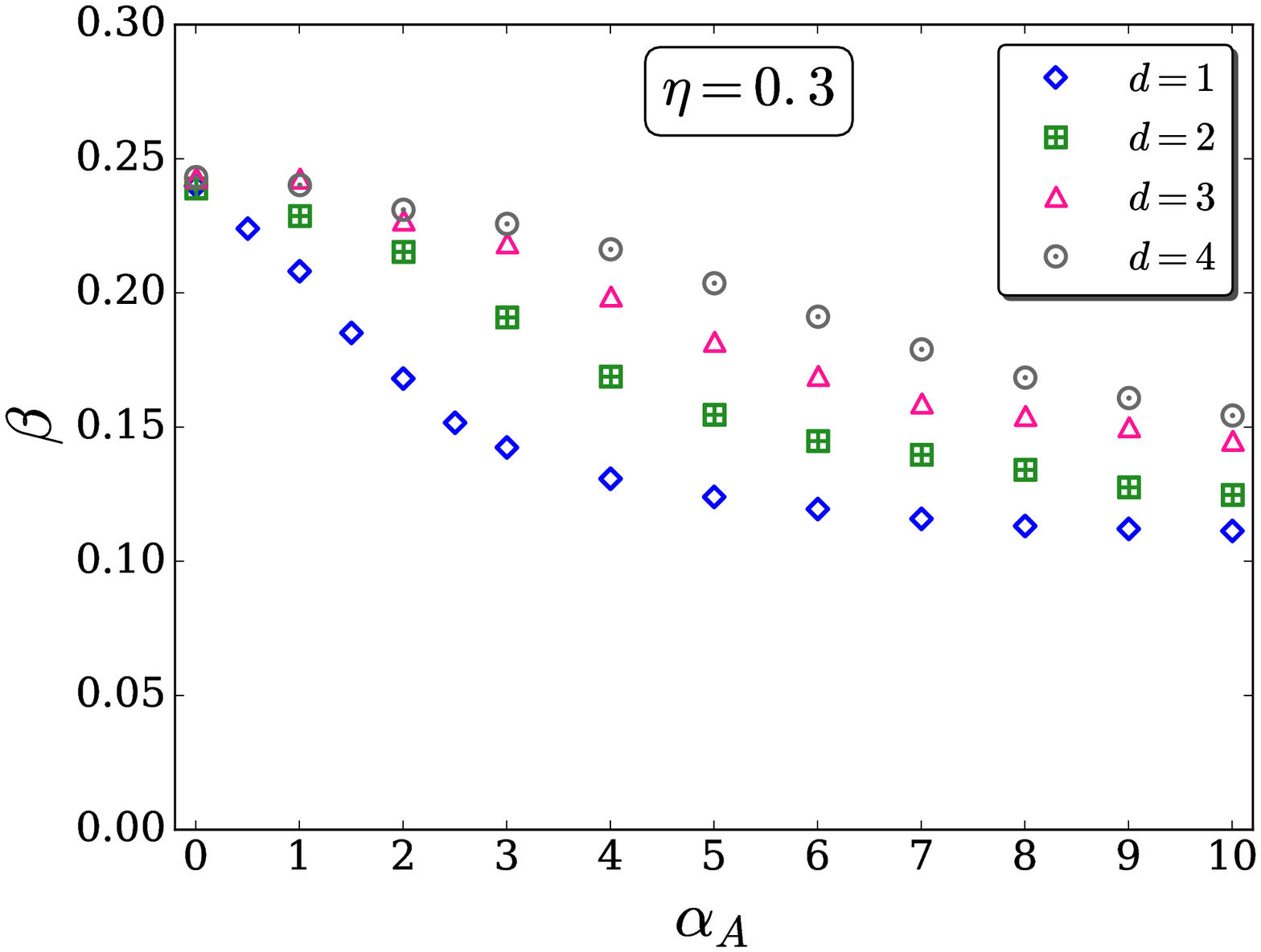}}
\subfigure {\includegraphics[scale = 0.37]{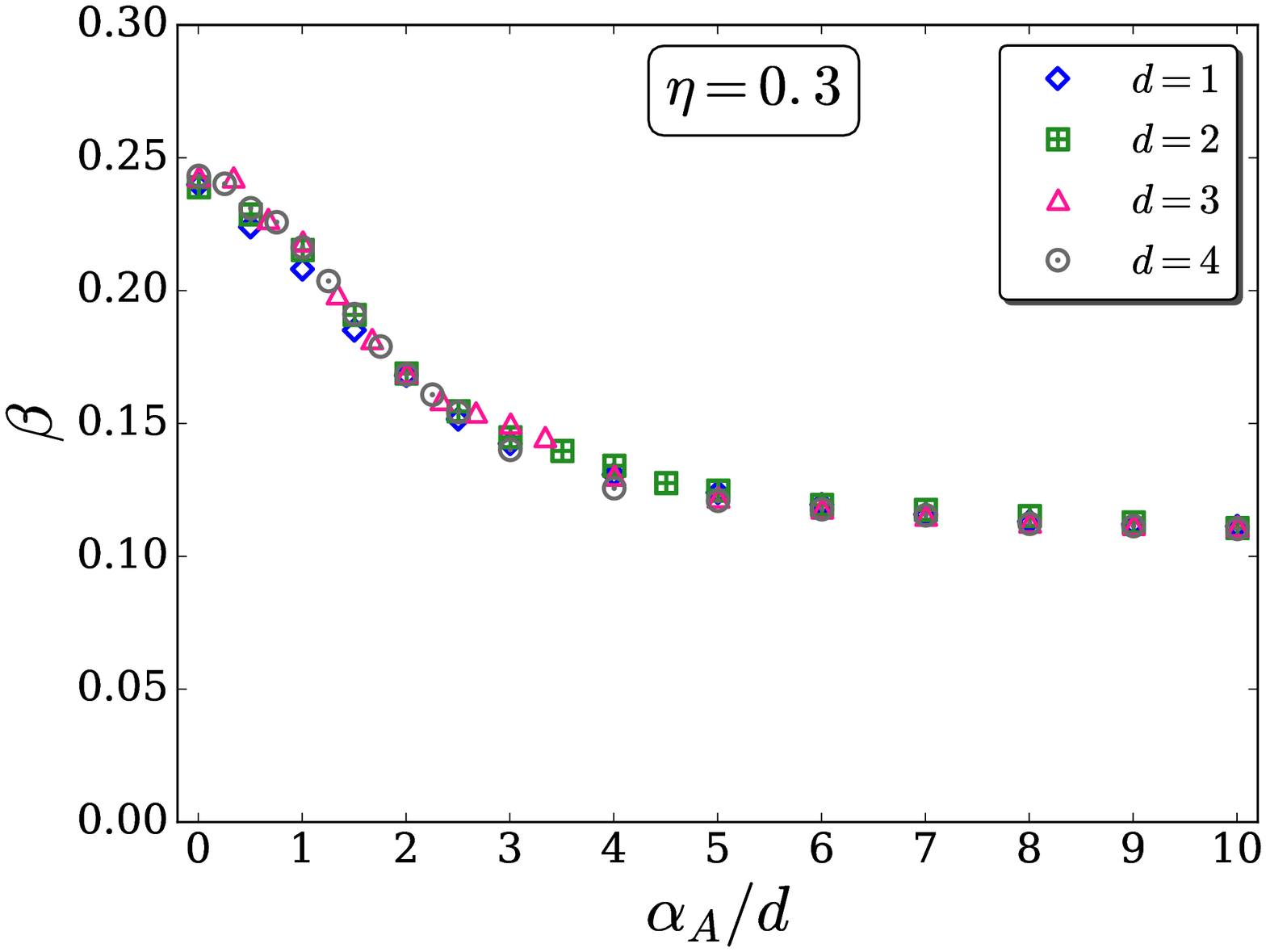}}
\subfigure {\includegraphics[scale = 0.37]{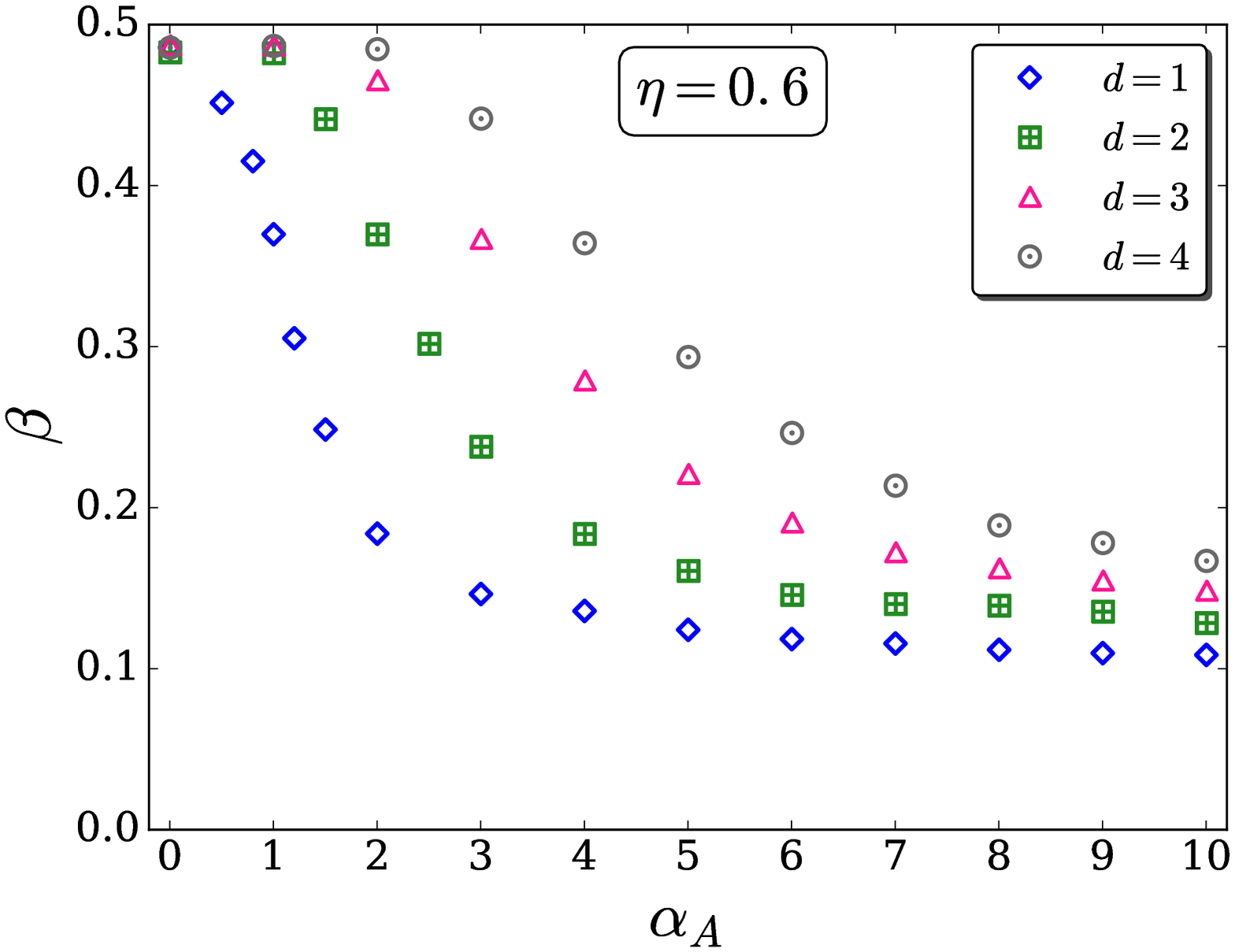}}
\subfigure {\includegraphics[scale = 0.37]{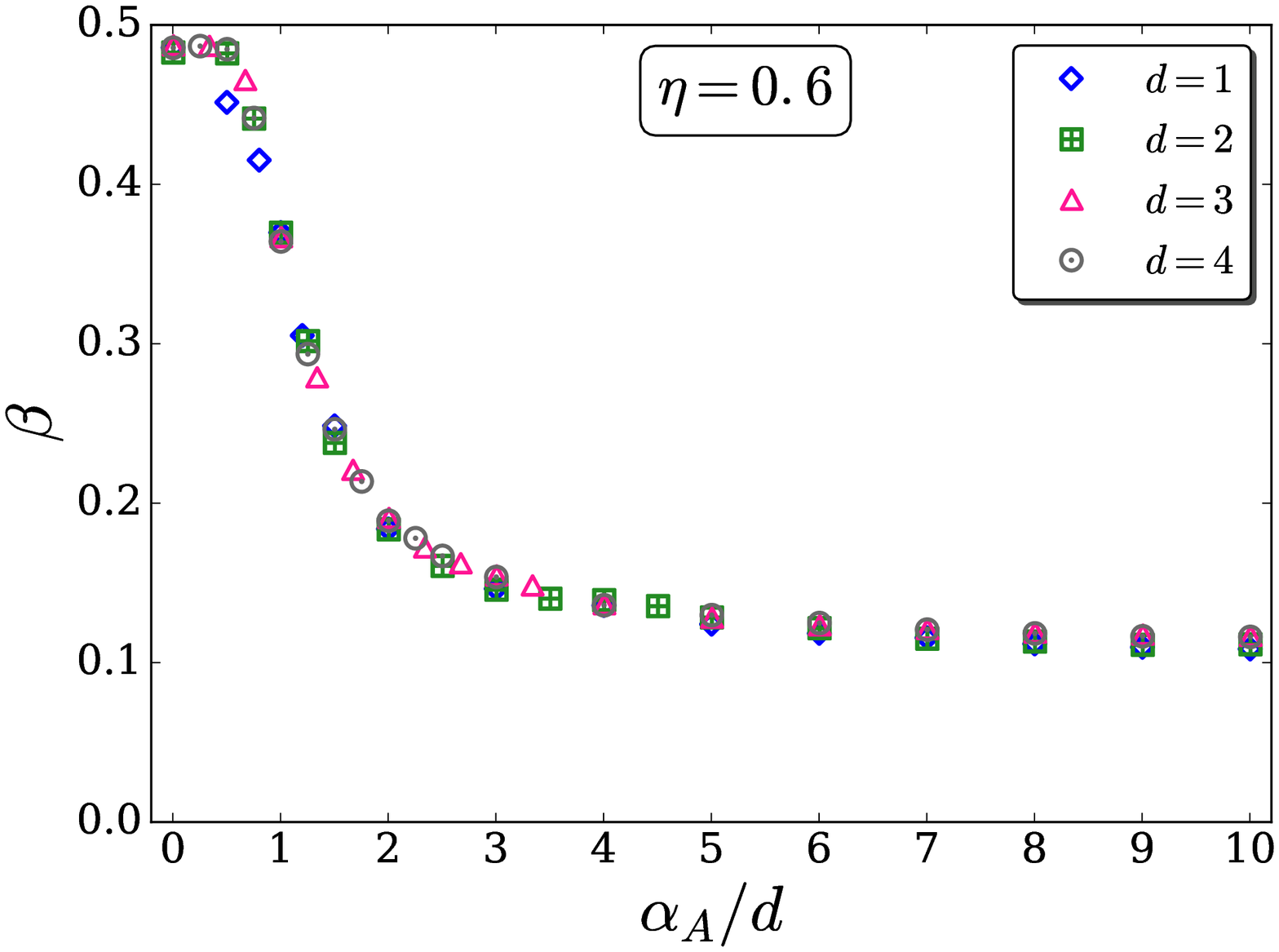}}
\subfigure {\includegraphics[scale = 0.37]{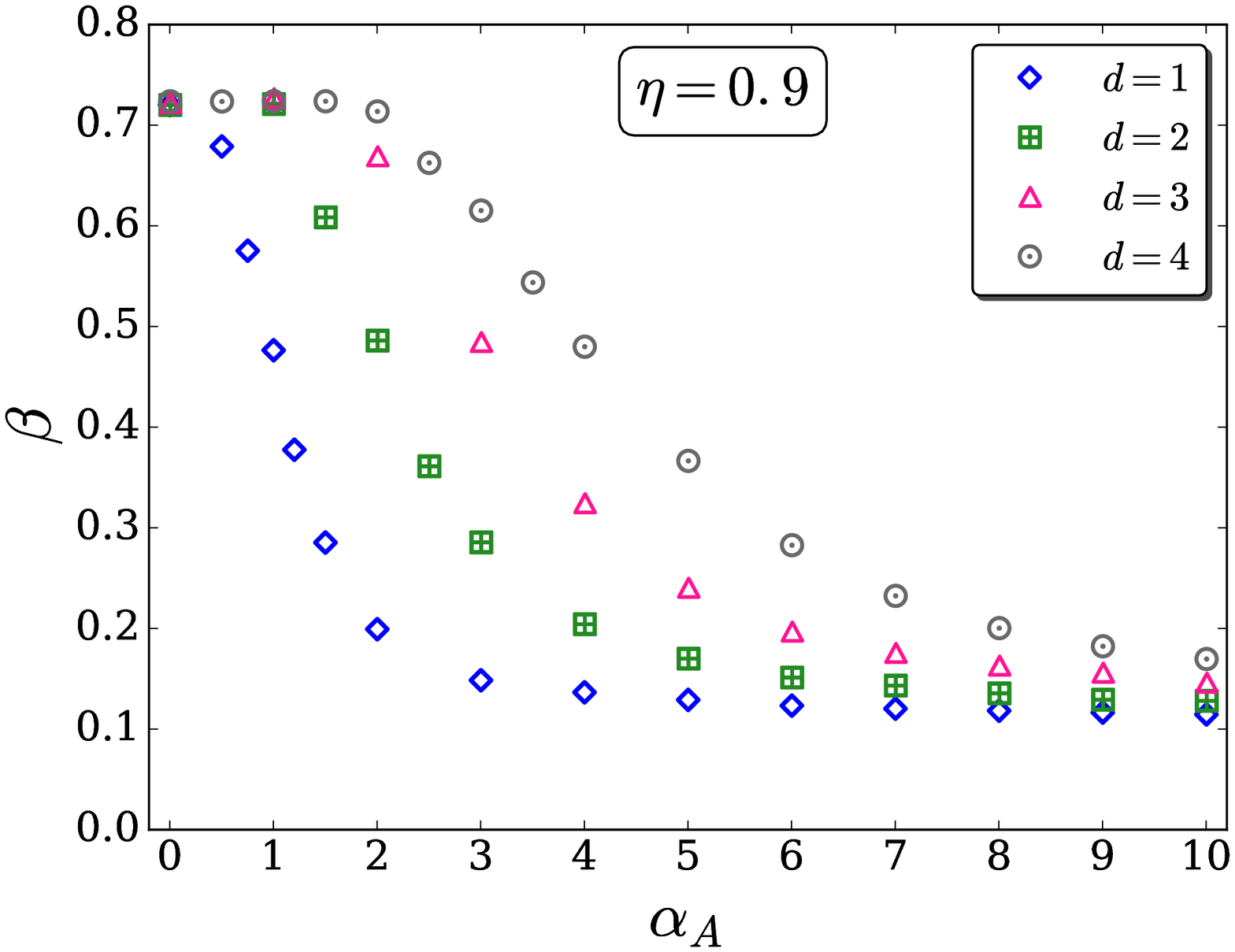}}
\subfigure {\includegraphics[scale = 0.37]{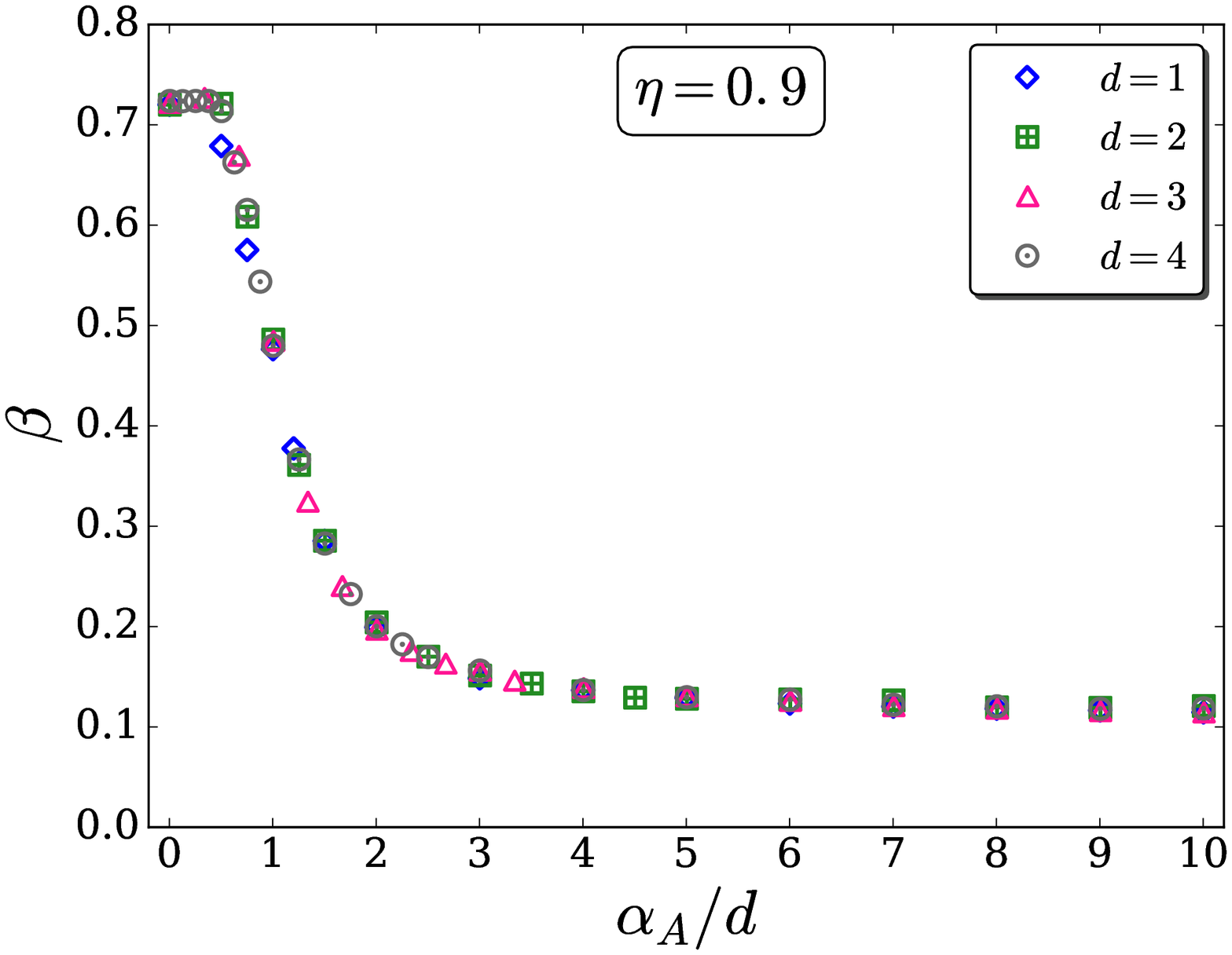}}
\end{center}
\caption{\small $\beta$ versus $\alpha_A$ graph for the site $i = 10$ with $\eta = 0.3$, $\eta = 0.6$ and $\eta = 0.9$. We can see that $\beta$ depends on $\alpha_A$ and $d$. In $\beta$ versus $\alpha_A/d$ graph we can see that this result exhibit the $\beta$ universality. However, we have different universal $\beta$ curves for different $\eta$ values. The simulations have been run for $10^{3}$ samples of $N = 10^{5}$ sites each.}
\label{fig8} 
\end{figure}

In the figure \ref{fig8} we show the exponent $\beta(\alpha_A, \eta_i, d)$ versus $\alpha_A$ and versus $\alpha_A/d$ for some typical values of $\eta$. To analyze the $\beta$ exponent, we fixed the value of $\eta$ for the site $i=10$ ($\eta$ is randomly chosen for all the sites excepting that one for which we want to compute the connectivity time evolution), and we observed how $\beta$ varies with  $\alpha_A$, $d$ and $\eta$. Our results show that $\beta$ depends of $\alpha_A$, $d$ and $\eta$. However, interestingly enough, for any fixed value of $\eta$, the $\beta$ curves have an universal behavior with respect to the scaled variable $\alpha_A/d$. 

\begin{figure}[H]
\begin{center}
\includegraphics[scale = 0.48]{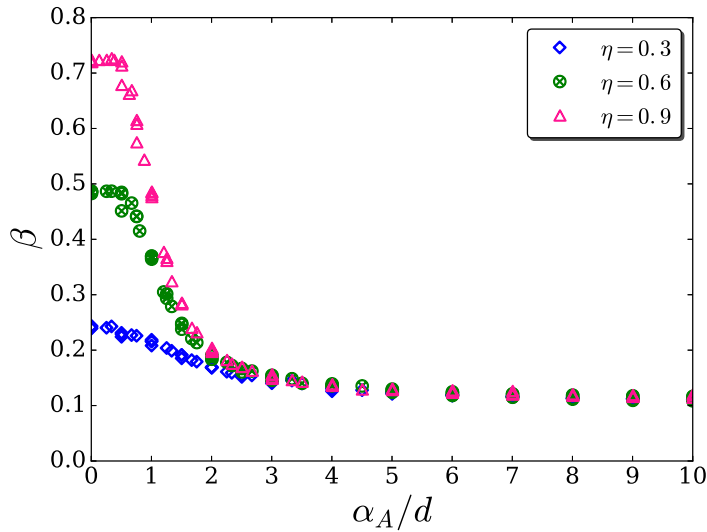}
\caption{\small Illustration of the $\beta$ versus $\alpha_A/d$ curves for  three different values of $\eta$ including all dimensions ($d=1,2,3,4$). For $\alpha_A/d = 0$ our results agree with the Bianconi-Barab\'asi model with $\beta \simeq 0.8\eta$; for $\alpha_A/d \to \infty$, the $\beta$ exponent appears to approach a constant value close to $0.1$ for all values of $\eta$.} 
\label{fig9}
\end{center}
\end{figure}

In the figure \ref{fig9} we can see the collapse of the $\beta$ curves versus $\alpha_A/d$ for different $\eta$ values. The values of the $\beta$ exponent for $\alpha_A/d \gtrsim 2$ show that the fitness parameter does not influence the connectivity time evolution of the sites. In this limit, it does not matter if the site has a large or small fitness, the rate at which it acquires links will be the same.


\section{Conclusion and discussion}
Along the lines of the paper \cite{britosilvatsallis2016}, we have introduced a $d$-dimensional growth model which, within the preferential attachment rule, includes connectivity, metrics, and fitness. The importance of the distance in the preferential attachment rule is more pronounced for increasing values of $\alpha_A$. We saw that the connectivity distribution and the dynamical $\beta$ exponent are substantially influenced by both $\alpha_A$ and $d$. We have shown that the degree distribution $P(k)$ is (numerically) very well fitted by the $q$-exponential function that naturally emerges from nonextensive statistics. When $0 < \alpha_A/d \leq 1$ all degree distributions are characterized by $q$-exponential functions, displaying in this limit the same universality with $q = 7/5$. Our most remarkable results show that $q$, $\kappa$ and $\beta$ present universal curves with respect to the scaled variable $\alpha_A/d$. Finally, we verified that there is a regime, $\alpha_A/d \gtrsim 2$, where the fitness parameter does not influence the connectivity time evolution of the sites in these networks. Summarizing, our results reveal a strong connection between (asymptotically) scale-free networks (with either short- or long-range interactions) and nonextensive statistical mechanics. Another important fact that our results exhibit is the intriguing ubiquity of the $\alpha_A/d$ variable, whose deep meaning in complex networks needs to be further explored.
\section*{Acknowledgments}
 We gratefully acknowledge partial financial support from CAPES, CNPq, Funpec, Faperj (Brazilian agencies) and from the John Templeton Foundation-USA.

\end{document}